\documentclass[referee]{raa}            
\usepackage{graphicx,times}             

\usepackage{color}
\usepackage{ulem}

\begin{document}
	
\title{The optimization of satellite orbit for Space-VLBI 
observation}{Optimization of satellite orbit for SVLBI obs.}

  \author{Lei Liu\inst{1,2,3}, Weimin Zheng\inst{1,2,3}}

\institute{$^1$Shanghai Astronomical Observatory, Chinese Academy of 
	Sciences; liulei@shao.ac.cn, zhwm@shao.ac.cn\\
	$^2$Key Laboratory of Radio Astronomy, Chinese Academy of Sciences, 
	Nanjing 210008, China\\
	$^3$Shanghai Key Laboratory of Space Navigation and Positioning 
	Techniques, Shanghai 200030, China}
   \date{}

\abstract{By sending one or more telescopes into space, Space-VLBI (SVLBI) is 
able to achieve even
higher angular resolution and is therefore the trend of the VLBI technique. For 
SVLBI program, the design of satellite orbits plays an important role for the 
success of planned observation. In this paper, we present our orbit 
optimization scheme, so as to facilitate the design of satellite orbit for 
SVLBI observation. To achieve that, we characterize the $uv$ coverage with 
a measure index and minimize it by finding out the corresponding orbit 
configuration. 
In this way, the design of satellite orbit is converted to an optimization 
problem. We can prove that, with appropriate global minimization method, the 
best orbit configuration can be found within the reasonable time. Besides that, 
we demonstrate this scheme can be used for the scheduling of 
SVLBI observations. }

\authorrunning{Lei Liu \& Weiming Zheng}
\titlerunning{The optimization of satellite orbit for Space-VLBI 
	observation}
\maketitle

\section{Introduction}\label{sec:intro}

VLBI (Very Long Baseline Interferometry) is the astronomical 
instrument with 
the highest angular resolution, and is therefore widely used in the field of 
astrophysics (Event Horizon Telescope Collaboration 2019), astrometry (Ma et 
al. 2009; Schuh \& 
Behrend 2012) and 
deep space exploration (Duev et al. 2012; Zheng et al. 2014). The resolution of 
VLBI is 
proportional to the baseline 
length and the observation frequency (Thompson et al. 
2001). For ground based VLBI, the 
length of a baseline is limited by the size of the Earth. E.g. the Event 
Horizon Telescope achieved an angular resolution of 23$\mu$as at 230~GHz with 
baseline length comparable to the Earth diameter (Event Horizon Telescope 
Collaboration 2019).
To achieve 
even higher angular resolution at the given frequency, one natural choice is to 
send one or more telescopes into space, which is the so called 
Space-VLBI (SVLBI; Gurvits 2018). 

Japan sent the first VLBI satellite VSOP (VLBI Space Observatory) into space in 
1997 (Hirabayashi et al. 1998, 2000). It was equipped with an 8.8 meter 
parabola 
antenna and 
works in 1.6 and 5~GHz. The orbit height was 22,000~km. The mission came 
to 
an end in 2005. Another SVLBI program was RadioAstron by Russia. It was 
launched in 2011 and worked until 2019. The designed observation frequencies 
were 0.3, 1.6, 5, 22~GHz. The orbit height was 338,000~km, which 
helped the 
project achieve the highest angular resolution of 7~$\mu$as at 22~GHz for SVLBI 
observations (Kardashev et al. 2012). 

Several SVLBI projects are under development in China (An et al. 2020). 
Shanghai 
Astronomical Observatory (SHAO) proposed the mission concept of space 
mm-wavelength VLBI array (SMVA) in 2010s (Hong et al. 2014). With the support 
of 
Chinese Academy of Sciences, prototype studies are conducted for the technical 
feasibility of the mission. One of the main achievement is the 10-m space 
antenna prototype (Hong et al. 2014; Liu et al. 2016). 

At present, SHAO is proposing a new SVLBI project: the Space Low 
Frequency 
Radio 
Observatory. In this 
project, two satellites each equipped with a 30 meter radio telescope will be 
sent into the Earth elliptical orbit (orbit height 2,000~km $\times$ 
90,000~km). The observation frequency ranges from 30~MHz to 1.7GHz. Two 
telescopes 
will conduct
collaborate observation with FAST (Five-hundred-meter Aperture Spherical radio 
Telescope), SKA (Square Kilometer Array) and other ground based 100~m level 
large radio 
telescopes, so as to achieve both high angular resolution and sensitivity. 
This project is special, as two satellites dedicated to SVLBI 
observations 
make it standout from VSOP and RadioAstron, of which only one satellite is 
deployed. Also it is different from Chang'E missions, for which the orbit is 
fixed and will not be adjusted for SVLBI observations. For the first time, the 
project will provide unprecedented flexibility for the design of satellite 
orbit dedicated to VLBI observations. 
For VLBI observation, one of the most important applications is to obtain 
the high angular resolution image of the target (radio imaging). In this 
process, a good $uv$ coverage is essential for obtaining an ideal antenna beam, 
and 
finally determines the quality of the image. 
However, this is not a trivial task. 
First of all, by looking through literature, one may realize that there are
no commonly accepted rules for the characterization of a ``good'' $uv$ 
coverage that is suitable for radio imaging. Moreover, a satellite orbit is 
uniquely described by 6 orbital elements. Orbit design for two 
satellites involves the combination of 12 such kind of elements. It is 
actually computationally challenging to find out the orbit configuration 
that 
yields the best ``uv'' coverage from the large parameter space. 

The design of satellite orbit for SVLBI observations is somewhat similar to 
the  classical array configuration problem that has been well studied in the 
last three decades. Although the trajectory of a space telescope is quite 
different from that of ground based telescopes, we can still 
gain inspirations from previous work. Keto (1997) propose the 
array shape of a curve with constant width, so as to achieve a uniform sampling 
in the $uv$ plane.
Boone (2001) argued that a Gaussian radial and uniform angular 
distribution in 
the $uv$ plane yielded a Gaussian shaped synthesis beam. Therefore, they 
optimized the array 
based on the discrepancy (the ``pressure force'') between the model and the 
actual coverage. Kogan (1997) took another approach by 
minimizing the side lobes. Su et al. (2004) optimize the array 
distribution by taking a 
``thieving'' approach. Karastergiou et al. (2006)
characterized the $uv$ coverage 
with a single quantity and go through all possible array configurations to 
minimize it. 

In this paper, we try to solve the orbit design problem by drawing 
inspirations from previous work and taking into account the orbit 
configuration.
In short, we characterize the $uv$ coverage with a measure index and minimize 
it 
by 
finding out the corresponding satellite orbit configuration. In this way, 
orbit
design is converted to an optimization problem. Our work proves that 
this approach is feasible, the best orbit configuration can be found within
a reasonable 
time using modern global minimization method. 

This paper is organized as follows: in Sec.~\ref{sec:scheme} we introduce the 
optimization scheme; in Sec.~\ref{sec:application} we 
present its application, including the design of orbit and the schedule of 
observation. In Sec.~\ref{sec:sum} we discuss the results and present 
summary and conclusions.

\section{The orbit design and optimization scheme}\label{sec:scheme}
\begin{figure}
	\centering
	\includegraphics[width=\columnwidth]{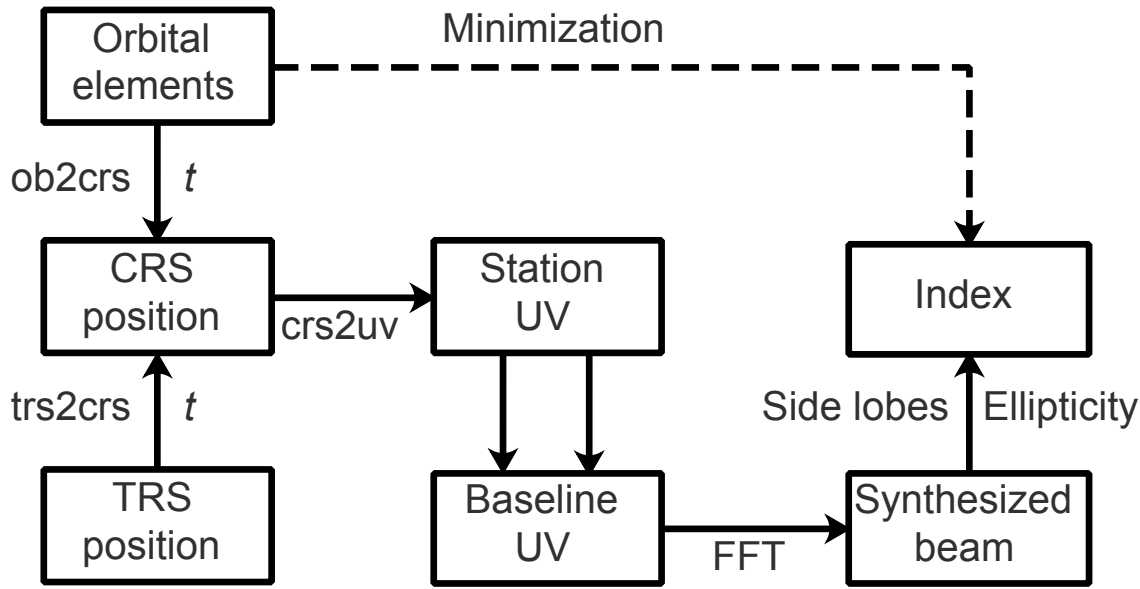}
	\caption{Data flow chart of the orbit design and optimization scheme.} 
	\label{fig:dataflow}
\end{figure}

In this work, we characterize the $uv$ coverage with an index $s=s(uv)$.
Given a specific time range, the $uv$ coverage is determined by the 
configuration of satellite orbit.
\footnote{It is actually the trajectories of satellites and ground stations 
together determine the $uv$ coverage. However, the latter part only depends 
on earth rotation and is fixed at given time range. The variation of $uv$ 
coverage is determined by orbit 
configurations.}  According to the classical satellite dynamics, the 
orbit of a given celestial 
object is described by 6 elements: semi-major axis $a$, eccentricity 
$e$, inclination $i$, right ascension of ascending node $\Omega$, argument of 
periapsis $\omega$ and mean anomaly at reference epoch $M_0$. Accordingly, 
for 
two satellites in the planned array there will be 12 such kind of elements. As 
a result, the index can 
be expressed as an optimization function of those elements. In this way, the 
design of satellite orbits is converted to an optimization problem. 

The data flow of the optimization scheme is demonstrated in 
Fig.~\ref{fig:dataflow}. For each given moment of time $t$, the telescope 
position (or state vector) is calculated in Celestial Reference System (CRS). 
For satellite, this routine is based on the orbital elements. For ground based
stations, transformation of station coordinates from TRS (Terrestrial 
Reference System) to CRS is conducted. In this process, we take into 
account the Earth rotation, precession, nutation and polar motion effects
\footnote{The precession, nutation and polar motion matrices require earth 
orientation parameters, which are routinely updated by IERS.}. Once the station 
CRS positions are obtained, we calculate their projections on the $uv$ plane, 
and further calculate the baseline $uv$. The next step is to characterize the 
$uv$ coverage with an index, which is based on the synthesized beam in the 
image plane. For radio imaging, a good $uv$ coverage yields a smooth 
synthesized beam, which is crucial for the quality of the final image.
The details on the index calculation are provided in the next section.
To this point, the 
index is actually a function of satellite orbital elements. Thus, it would 
be possible to find the best orbit configuration by minimizing the index 
value. 

\begin{figure*}[t]
	\centering
	\includegraphics[width=\columnwidth]{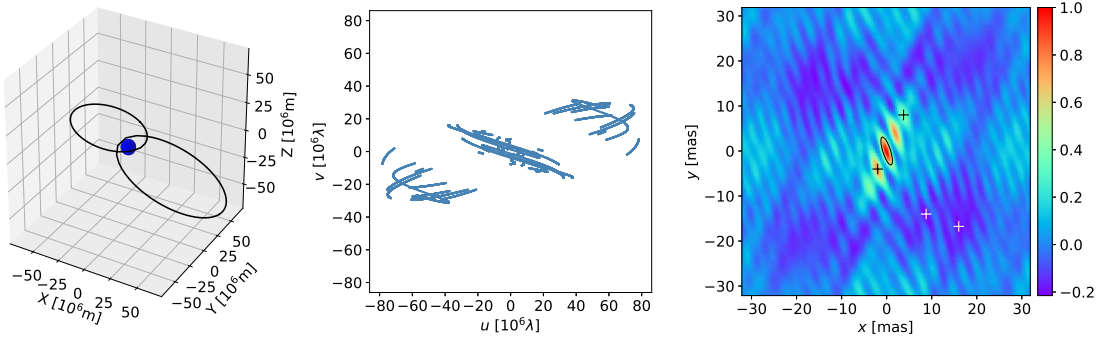}
	\caption{Orbit configuration (left), $uv$ coverage (middle) and beam 
		pattern (right) before optimization. The main lobe of the beam is 
		shown 
		with a black 
		ellipse. Orbital elements are selected randomly by the optimization 
		function. The first and second peaks and nadirs of the side lobe are 
		marked 
		with black and white crosses, respectively. 
		Index: 1.718193 (L1), semi-major axis: 52,378.1~km,
		eccentricity: 0.84, inclination: -21.4$^\circ$/6.3$^\circ$,
		right ascension of ascending node: -134.3$^\circ$/-129.3$^\circ$,
		argument of periapsis: -51.2$^\circ$/92.4$^\circ$,
		mean anomaly: 163.5$^\circ$/38.9$^\circ$.}\label{fig:uvbeam_raw}
\end{figure*}

\subsection{Characterization of $uv$ coverage}

\begin{figure*}[t]
	\centering
	\includegraphics[width=\columnwidth]{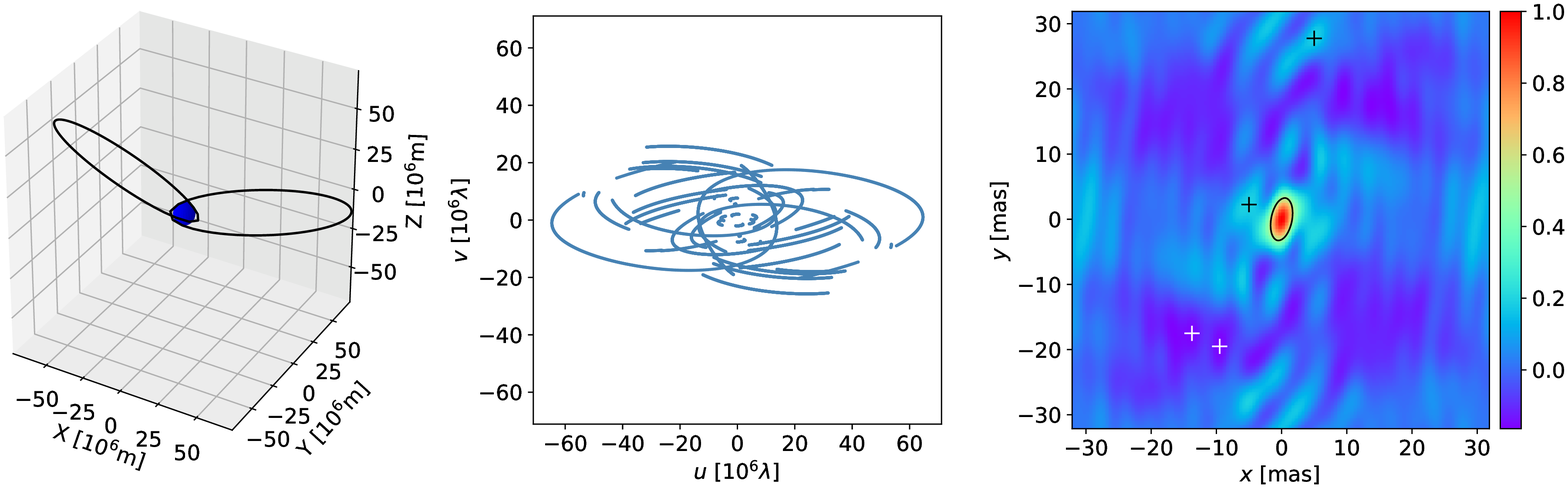}
	\caption{Orbit configuration (left), $uv$ coverage (middle) and beam 
		pattern (right) after optimization with L1 norm. Descriptions of 
		ellipse and 
		crosses are provided in the caption of 
		Fig.~\ref{fig:uvbeam_raw}. 
		Index: 0.749449 (L1), semi-major axis: 52,378.1~km,
		eccentricity: 0.84, inclination: 151.5$^\circ$/-4.6$^\circ$,
		right ascension of ascending node: -41.3$^\circ$/45.9$^\circ$,
		argument of periapsis: -38.8$^\circ$/163.6$^\circ$,
		mean anomaly: -19.5$^\circ$/-91.0$^\circ$.
	}\label{fig:uvbeam_final}
\end{figure*}

\begin{figure*}[t]
	\centering
	\includegraphics[width=\columnwidth]{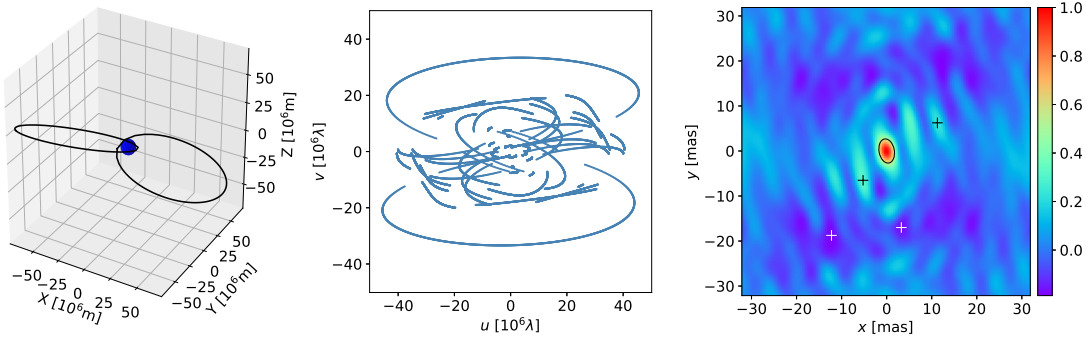}
	\caption{Orbit configuration (left), $uv$ coverage (middle) and beam 
		pattern (right) after optimization with L2 norm. Meanings of ellipse and
		crosses are explained in the caption of 	
		Fig.~\ref{fig:uvbeam_raw}.
		Index: 0.488062 (L2), semi-major axis: 52,378.1~km,
		eccentricity: 0.84, inclination: -163.1$^\circ$/-162.5$^\circ$,
		right ascension of ascending node: -177.7$^\circ$/4.8$^\circ$,
		argument of periapsis: 159.3$^\circ$/-167.1$^\circ$,
		mean anomaly: 153.5$^\circ$/-2.1$^\circ$.	
	}\label{fig:uvbeam_l2}
\end{figure*}

The key idea of this work is to find a quantity that characterizes the 
$uv$ 
coverage appropriately. In our view, it must fulfill the following 
requirements:
\begin{itemize}
\item Scalar form. Suitable for minimization.
\item Accurate. Smaller value yields better coverage.
\item Easy to calculate. Since the calculation will be conducted many times 
when investigating the huge parameter space, calculation speed is very 
important.
\end{itemize}
We draw inspirations from previous work. Initially, 
we took the idea of Boone (2001) and used an index to evaluate 
the 
discrepancy between a Gaussian shaped radial distribution and actual data. Soon 
we realized that the Gaussian distribution was difficult to achieve for SVLBI 
observations of which the $uv$ coverage is usually sparse. Actually this is 
already pointed out by Karastergiou et al. (2006) that Boone 
(2001) scheme is mainly for dense 
interferometer. Eventually we decided to characterize the $uv$ coverage 
with 
the synthesized beam in the image plane. Although the $uv$ coverage and the 
synthesized beam are mathematically equivalent, the latter one is relatively 
easier to evaluate: an ideal beam should be smooth and round (less fluctuate 
and oblate) in shape. 
We propose a measure index of the following form:
\begin{equation}
s_\mathrm{L1} = w_r r_\mathrm{L1} + w_e e. 
\end{equation}
Here $r_\mathrm{L1}=(a_1 + a_2 + |a_{-1}| + |a_{-2}|)/a_0$ is the ratio of side 
lobes 
to the main lobe, which measures the fluctuation of the beam. $a_i$ represent 
the value of the $i$-th peak/nadir of the beam pattern. $a_0$ is the main 
lobe. 
$a_1$ and $a_2$ are the first and the second side lobes. $a_{-1}$ and 
$a_{-2}$ are 
the first and the second nadirs, which are negative. $e = 
b_\mathrm{maj} / b_\mathrm{min}-1$ measures the ellipticity of the beam.  
$b_\mathrm{maj}$ and $b_\mathrm{min}$ are the major and minor axis of the beam, 
which are derived from the $uv$ coverage with TPJ's algorithm in 
DIFMAP (Shepherd 1997). $w_r$ and $w_e$ are the weights of the two terms. 
In this work, we set them to 0.9 and 0.1, respectively. Initially the 
ellipticity of the beam is not taken into account. Soon 
we realize that this might lead to an extremely oblate beam (large 
ellipticity). According to our 
test, a weight of 0.1 for the ellipticity term effectively reduces the 
oblateness of the beam in the optimization process. We have to point out that 
our choice of 
weights is somewhat 
arbitrary. Their values could be further adjusted to achieve a better 
optimization result in the actual application. 
Our optimization approach is similar with Kogan 
(1997), which 
adjust 
antenna positions to reduce the side lobes. One may find that the sum of 
absolute values correspond to the L1 norm. It is also worth investigating the 
L2 norm:
\begin{equation}
s_\mathrm{L2} = w_r r_\mathrm{L2} + w_e e. 
\end{equation}
Here $r_\mathrm{L2}=(a_1^2 + a_2^2 + a_{-1}^2 + a_{-2}^2)^{1/2}/a_0$. 

\subsection{Constrains of Space-VLBI observation}\label{sec:constraint}
The space low frequency array project is still in its preliminary stage, which  
gives us a lot of  
flexibility to design the orbit. However, VLBI is a complex technique,  
placing one or more antennas in the Earth orbit introduces many extra 
uncertainties and makes orbit design even more difficult. As a result, there 
are still 
some constraints that must be taken into account. We list them below and 
discuss 
their influence on orbit design.

\begin{itemize}
\item Orbit height. According to the preliminary plan, the VLBI satellite 
will be 
sent into the Earth elliptical orbit by rocket. In principle, the 
rocket itself has no special requirement for the orbit. For VLBI consideration,
the apogee height is set to 90,000~km, such that the baseline length is one 
order magnitude longer than that of the ground based VLBI. Meanwhile, the 
perigee height is fixed at 2,000~km, so as to guarantee a data transmission 
rate of 
1.5~Gbps in X or Ka band. These constraints fix the semi-major axis and the 
eccentricity, which reduce the parameters from 12 to 8 and 
therefore speedup the optimization process.

\item Observation time. Having the orbit height such as mentioned above, 
the corresponding orbit period will be 33.1 hours. To obtain a good $uv$ 
coverage, observations to 
 the target source should cover the whole elliptical orbit. However, it is not 
 realistic to require that the observation is continuous in the whole orbit 
 period. As a result, observations 
 should be conducted several times when the satellite is located at 
 different parts of the orbit.
 

\item Collaboration with ground based telescopes.
When conducting VLBI observations, two satellites will collaborate with the 
ground based large telescopes, so as to achieve both 
high angular resolution and sensitivity simultaneously. When simulating 
$uv$ coverage, contributions from ground-ground, space-ground baselines 
must be 
taken into account and will determine the final beam patterns together with
space-space baselines. As a result, their presence will affect on 
the selection of orbital elements in the optimization process.
\end{itemize}

\subsection{Implementation of optimization scheme}

Concerning the complex relationship between satellite orbit and the 
corresponding 
$uv$ coverage, the optimization function cannot be described analytically. 
Moreover, it is not guaranteed that the function is convex, which means there 
might be many local minima that must be avoided when looking for global 
minima. 
Optimization is a large topic in applied mathematics. It is completely not our 
intention to develop an optimization method from scratch for our work. 
Fortunately, there are many well developed global minimization methods 
which
have been implemented in Python scipy package. Among them we 
choose the ``differential\_evolution'' method (Storn \& Price 
1997). 
According to our test, it is able to find out the 
solution (global minimum) within the reasonable time(around 15~min for each 
solution), as we will describe in detail in Sec.~\ref{sec:opt_result}. We 
have to point out that this does not necessarily mean it 
outperforms other global optimization methods in the algorithmic level. 
According to our 
analysis, the main reason it converges faster than other methods is it 
provides parallel implementation in current scipy package.

\section{Applications}\label{sec:application}

\begin{figure}
	\centering
	\includegraphics[width=\columnwidth]{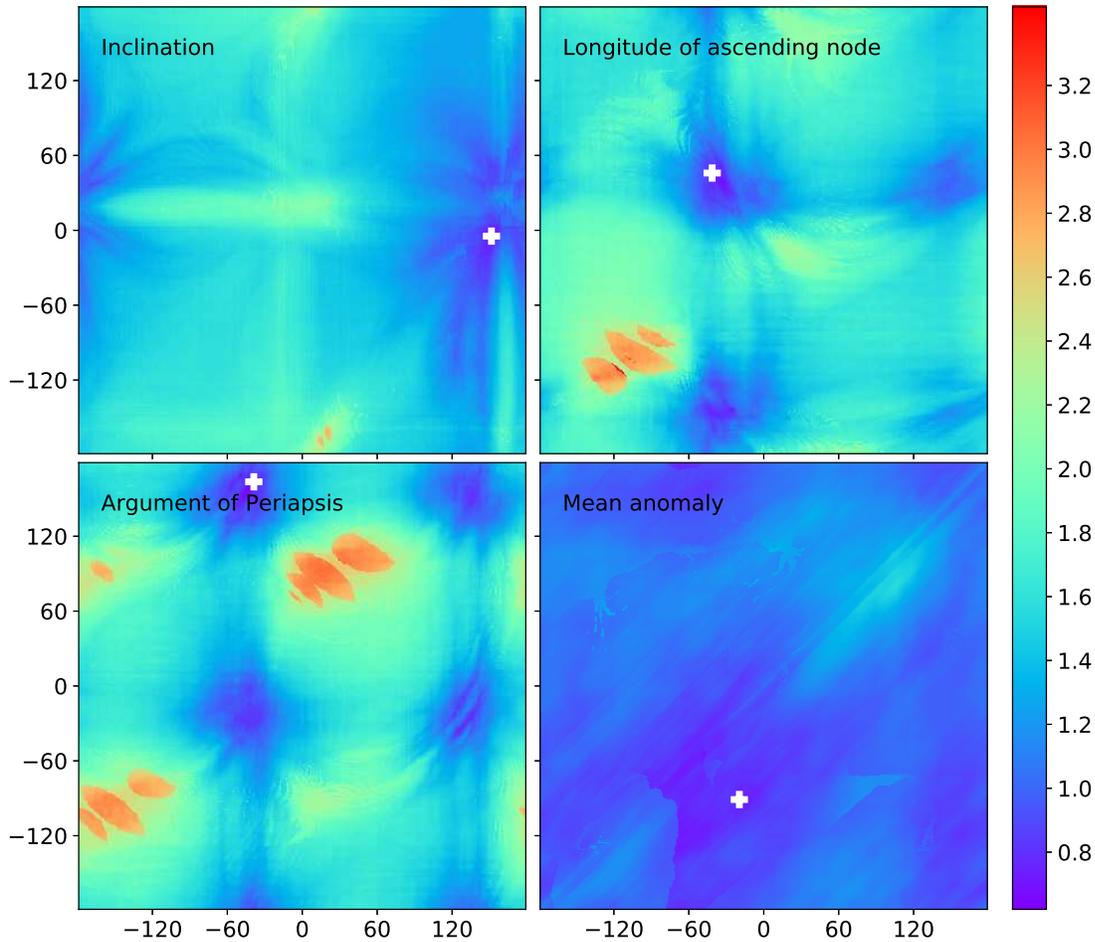}
	\caption{Snapshots of the optimization function in the phase space. 
	Index values at each pixel are calculated by investigating the 2 parameters 
	labeled on the top left of the corresponding panel while keeping other 6 
	parameters set to the L1 solution (see Fig.~\ref{fig:uvbeam_final} for the 
	value). The pixel size of the 4 panels is set to 1$^\circ$. White 
	cross in 
	each panel labels out the solution found by the 
	differential\_evolution method. } \label{fig:pmap}
\end{figure}

In this section, we present two applications of the optimization scheme: 
orbit design and observation scheduling, as described below.
 
\subsection{Orbital design}

\subsubsection{Observation setup}
We setup an observation for the application of the orbit design scheme. The 
assumed 
observation starts on 2020-03-11T00:00:00 
UTC and lasts for 24 hours. The target source is M87 (Park et al. 2019). The 
observation 
frequency 
is set to 300~MHz. According to the 
preliminary plan, two VLBI satellites take part in the observation. Five ground 
based 
telescopes, FAST (Five-hundred-meter Aperture Spherical radio 
Telescope; Nan et al. 2011), QTT (QiTai radio Telescope), Effelsberg 100~m adio 
telescope in Germany, GBT (Green Bank Telescope) in US and the planned SKA-low 
in Australia take part in the observation. FAST has conducted VLBI observations 
successfully 
with TMRT (TianMa Radio Telescope in Shanghai) last year. We are expecting more 
scientific breakthrough with its extremely high sensitivity. QTT is still in 
its construction stage, which is promising in conducting collaborated 
observation in the next 10 years. 
The minimal elevation angle is set to 30$^\circ$ for FAST, and 15$^\circ$ 
for 
other ground based telescopes. Besides that, to avoid radio interference from 
the Earth, we set a minimum separation angle of 5$^\circ$ between the 
source 
and the Earth surface at the satellite. 

\subsubsection{Optimization result}\label{sec:opt_result}
We use the ``differential\_evolution'' method provided by scipy 
package to find out the global minimum of the optimization function. 
The calculation of $uv$ coverage and the localization of 
zenith/nadir point of the beam pattern in each set of orbit configuration is a 
time consuming process. To keep a reasonable optimization speed, positions of 
satellites and ground based telescopes are sampled every 1 minute. For the 
beam pattern, the cell size is set to 0.25 mas, which is around 1/10 of the 
angular resolution for a 100,000~km baseline at 300~MHz.
The apogee and perigee heights are set to 90,000~km and 2,000~km, 
respectively. This fixes the semi-major axis and the eccentricity of the orbit. 
As a result, the optimization is conducted in the 8 dimensional parameter space 
composed of inclination, longitude of ascending node, argument of Periapsis 
and mean anomaly of the two satellites. Based on our 
actual implementation, it takes about 15 minutes on 12 workers (processes) in a 
server equipped with 4 Intel Xeon E7530 @ 1.87 GHz (24 physical cores in total) 
for each optimization in the paper. We have to emphasize that this result is 
strongly depends on the hardware platform.

Fig.~\ref{fig:uvbeam_raw}, \ref{fig:uvbeam_final} and \ref{fig:uvbeam_l2} 
demonstrate the orbit, $uv$ coverage and beam pattern before and after 
optimization. The orbital
elements (solutions) are presented in the caption of the corresponding 
figures. By observing the main lobe and the surrounding region, we may find 
that 
the 
optimized result is much more smooth and less oblate. This is consistent with 
our design of the optimization function. The orbit configuration selected by 
the  
optimization routine yields the smallest index. We may expect that compared 
with the unoptimized beam, the optimized Gaussian shaped beam is more suitable 
for deconvolution. Also note that the minor axis of the beam pattern 
corresponds to an elongated $uv$ distribution in the same direction, and vice 
verse. This is consistent with the radio imaging theory: the angular resolution 
is proportional to the inverse of the baseline length ($\theta\sim\lambda/d$).
We have to point that although the $uv$ coverage of 
the optimized orbit is much better than that of the unoptimized, the baseline 
length of the optimized orbit is shorter, which means we obtain a good beam 
pattern at the expense of lower 
angular resolution. However, we think this is worthwhile since a small beam 
with 
large fluctuation is of no help for imaging. We have to point out that 
such kind of long baselines might be necessary in some non-imaging 
applications, e.g., high angular resolution astrometry. If we care
only about the position and the target is a point source, a little bit 
worse beam pattern is not a problem and can be overcome (Liu et al. 2019). 

\subsubsection{The complex structure of phase space}
Fig.~\ref{fig:pmap} present the snapshots of the optimization function in the 
phase space. The purpose of these snapshots is to demonstrate the complex 
structures of the parameter phase space. The 8 parameters are tangled together 
in a highly non-linear way, which makes the localization of the minimum 
point a difficult task: the global optimization method should avoid getting 
trapped in the ``local minimum''. Our work proves that with an appropriate 
method, 
the global minimum solution can be found within the reasonable time. 

\subsection{Observation scheduling}

\begin{figure}
	\centering
	\includegraphics[width=\columnwidth]{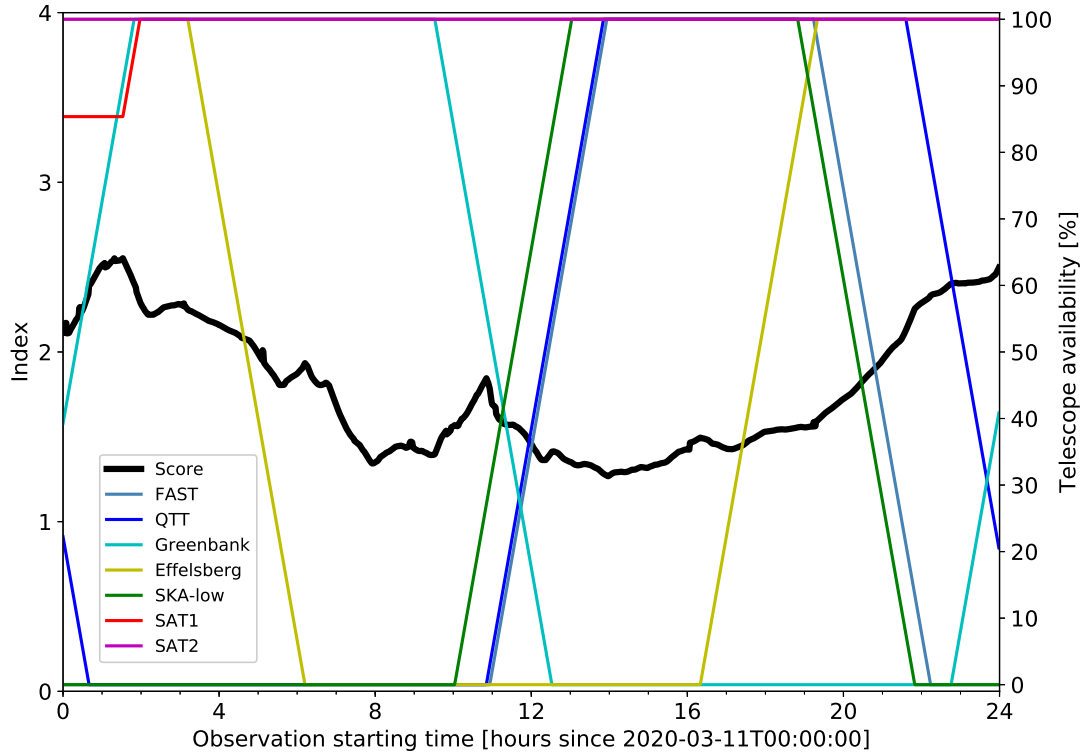}
	\caption{The index as a function of observation starting time. The planned 
	observation is a 3 hours session for M87. Thick black line and thin colored 
	lines correspond to the index and telescope availabilities, respectively.} 
	\label{fig:sched}
\end{figure}

\subsubsection{Motivation}
Another possible application of the optimization scheme is observation 
scheduling. There are already mature VLBI schedule programs for ground based 
telescopes. E.g., ``Sked'' (Gipson 2010) and ``VieSched++'' 
(Schartner \& B{\"o}hm 2019) for geodetic observations, 
``Sched''\footnote{http://www.aoc.nrao.edu/software/sched/} for 
astrophysical observations. However, the scheduling of SVLBI observations is 
still a blank area. For SVLBI, the position of VLBI satellite is determined by 
the orbit configuration instead of the Earth rotation. Besides that, the 
calculation 
of telescope availability is quite different from that of ground based 
telescopes. The eclipse of the Earth, data storage and many other space 
related ingredients must be taken into account. Moreover, the $uv$ coverage
of SVLBI observations is usually poor. More attention should be paid to 
the resulting beam. All 
of these make the schedule of SVLBI observations quite different from that of 
ground based observations. As a result, mature schedule methods and 
programs cannot be used directly in SVLBI observations. It is very necessary to 
develop a new method that takes the features of SVLBI observations into 
account. 

\subsubsection{Scheduling result}
We have found that our characterization of the $uv$ coverage and 
optimization 
scheme is very suitable for the scheduling of SVLBI observation. E.g., given 
some time for a source, what is the best time range to conduct 
the observation? Our answer is the scheduling can be converted to an 
optimization problem. This is demonstrated in Fig.~\ref{fig:sched}, for a 3 
hours observation, the index that characterizes the $uv$ coverage is a 
function of observation starting time: the observation yields the small index 
if it 
starts at 14:00 when 2 satellites and 3 ground based telescopes are fully 
available for observation. This is consistent with a basic scheduling 
principle: to utilize as many telescopes in the observation as possible. 
All of above suggest 
that our characterization of the $uv$ coverage is very helpful in 
SVLBI scheduling.

\section{Conclusions and discussions}\label{sec:sum}
In this paper, we present our orbit optimization scheme for the design of VLBI 
satellite orbit. In this 
scheme, we characterize the $uv$ coverage with an index and minimize it by 
finding out the corresponding orbit configuration. In this way, the design of 
satellite orbit is converted to an optimization problem. To valid the scheme, 
we setup a 24 hours observation for M87. Five large ground based telescopes 
FAST, 
QTT, Greenbank, Effelsberg, SKA-low and two VLBI satellites take part in the 
observation. Although the 
structure of the optimization parameter phase space is complex, we can prove 
that with modern global minimization method, it is possible to find out the 
best orbit configuration within the reasonable time. Moreover, we demonstrate 
that our characterization of $uv$ coverage can be used for the scheduling 
of SVLBI observations. 

We want to point out that the optimization scheme could and should be  
improved continuously. First of all, current optimization function (index) is 
based on our understanding of a good $uv$ coverage, which deals only with
the fluctuation and the ellipticity of the beam pattern. However, in 
actual application more parameters should be taken into account, e.g., angular 
resolution, sensitivity, etc. All of these would contribute to the optimization 
function with appropriate weights. Besides that, in this work, we only 
demonstrate the optimization for one source. For an actual scientific project, 
the 
optimization function should be the combination of a list of target sources. As 
long as we have obtained adequate computational resources, it is not difficult 
for our 
scheme to take multiple sources into account. Moreover, for a real satellite, 
there will be definitely more constraints on the orbit configuration. E.g., 
orbit height, inclination, etc. Our scheme provides enough flexibility 
to include these constraints in the optimization process.

The design of satellite orbit for SVLBI observation is a blank area. We still 
have a long way to go to obtain a commonly accepted ``good'' orbit 
configuration. We hope our work is helpful for China's future SVLBI project. 

\begin{acknowledgements}This work is supported by the National Natural Science 
Foundation of China (Grant Nos. 11903067, 11973011, 11573057, U1831137, 
11703070), 
Shanghai Outstanding 
Academic Leaders Plan, the Strategic Priority Research Program of the Chinese 
Academy of Sciences, grant No. XDB23010200, Shanghai Key Laboratory of Space 
Navigation and Positioning Techniques (ZZXT-201902).
We would like to show our deep appreciation to Dr. Wu Jiang, Dr. Jongho Park, 
Ms. Yidan Huang, Mr. Xingfu Liu for their kind support in writing this paper.  
We thank the referee for suggestions that greatly helped to improve the quality 
of this paper. 
\end{acknowledgements}

\end{document}